\documentclass[11pt]{article}

\usepackage{graphicx}
\usepackage{amsmath,amsthm}
\usepackage{amssymb,latexsym}
\usepackage{enumerate}
\usepackage[shortlabels]{enumitem}
\usepackage{epigraph}
\usepackage[authoryear]{natbib}
\usepackage{mathrsfs}
\usepackage{verbatim}
\usepackage{mathabx}
\usepackage{relsize}
\usepackage{mathtools}
\usepackage{accents}
\usepackage{bm}
\usepackage[titletoc,title]{appendix}
\usepackage{url}
\usepackage{setspace}
\usepackage{color}
\usepackage{fixltx2e}
\usepackage[letterpaper, margin=1in]{geometry}
\makeatletter
\g@addto@macro\bfseries{\boldmath}
\makeatother

\newtheorem{thm}{Theorem}
\newtheorem{cor}{Corollary}[thm]
\newtheorem{lem}{Lemma}
\newtheorem{prop}{Proposition}[thm]

\theoremstyle{definition}

\theoremstyle{remark}

\newtheorem{exmp}{Example}[section]

\usepackage{geometry}
\usepackage{doc}
\usepackage{epstopdf}
\DeclarePairedDelimiter{\floor}{\lfloor}{\rfloor}

\newcommand\B{\rule[-1.2ex]{0pt}{0pt}} % Bottom strut

\begin{document}

% Add the title section.
%\maketitle
\begin{titlepage}
  \centering
%  \includegraphics[width=0.15\textwidth]{example-image-1x1}\par\vspace{1cm}
%  {\scshape\LARGE Columbidae University \par}
%  \vspace{1cm}
  {\scshape\Large {A Method for Winning at Lotteries}\par}
  \vspace{1.5cm}
  {\scshape\Large July 7, 2017\par}
  \vspace{2cm}
  {\Large Steven D. Moffitt\textsuperscript{\textdagger} and William T. Ziemba\textsuperscript{\textdaggerdbl}\par}

% Add an abstract.
\abstract{
\noindent We report a new result on lotteries --- that a well-funded syndicate has a purely mechanical strategy to achieve expected returns of 10\% to 25\% in an equiprobable lottery with no take and no carryover pool. We prove that an optimal strategy (Nash equilibrium) in a game between the syndicate and other players consists of betting one of each ticket (the ``trump ticket''), and extend that result to proportional ticket selection in non-equiprobable lotteries. The strategy can be adjusted to accommodate lottery taxes and carryover pools. No ``irrationality'' need be involved for the strategy to succeed --- it requires only that a large group of non-syndicate bettors each choose a few tickets independently. 
}

  \vfill
  \textdagger Adjunct Professor of Finance, Stuart School of Business, Illinois Institute of Technology and 
              Principal, Market Pattern Research, Inc.\par
  \textdaggerdbl Alumni Professor of Financial Modeling and Stochastic Optimization (Emeritus),
  University of British Columbia, Vancouver, BC, and Distinguished Visiting Research Associate, Systemic Risk Centre,
  London School of Economics, UK
\end{titlepage}

%**********************************************************************************************
%
% Body text.
%
%************************************************************************************************
\section{Introduction} \label{introduction}

We show that a group of individuals who coordinate their betting has a strategy to obtain positive expected gains in a ``fair'' lottery. To illustrate, consider a $1,000$ number lottery in which a ``crowd'' of one thousand individuals each purchase a \$1 quick pick (sampling with replacement) and a coordinating ``syndicate'' that acts as a single bettor and purchases one of each ticket combination for a total of \$1,000. On average the crowd bets nothing on $36.8\%$ of the tickets, has one combination on $36.8\%$ and two or more on the remaining ticket combinations. Since the syndicate always has exactly one winning ticket, its expected payoff can be calculated as a function of the number of the tickets held by the crowd. When the crowd has no winning ticket, the syndicate wins the entire jackpot of \$2,000. When the crowd has one winning ticket, the syndicate wins \$1,000 = \$2,000/2, and so on. Thus the syndicate's expected payoff is
\begin{align*}
    &\$2,000*0.368 + \$1,000*0.368 + \text{(additional terms for crowd holdings of 2 or more tickets)}\\
  > &\$1,104,
\end{align*}
where the first term is the contribution to the expected payoff when the crowd has no winning ticket, and the second when the crowd has exactly one winning ticket. Therefore the syndicate's expected return is positive even without terms involving the crowd's holding of 2 or more winners!

This paper has three parts. First, the expected returns are calculated for a simple equiprobable $1,000$ number lottery that has no take.\footnote{The \emph{take} is the fractional amount a lottery deducts from the betting pool.} We show that a syndicate that bets $1,000$ different tickets earns on average a $26.41\%$ return against a crowd of $1,000$ small players who independently bet one ticket each. A syndicate strategy of buying $n < 1,000$ different tickets is not profitable unless $n \ge 583$. 

Then, results are proven for equiprobable lotteries with $t$ tickets and no take. If the crowd bets $\$c$ and a syndicate bets $\$s$, it is demonstrated that the syndicate has a positive expected return if $s > (1 + cy)/(1 - y)$, where $y = (1 - t^{-1})^{c + 1}$ and it bets its tickets on a set which is as evenly distributed as possible. If $s \le t$ and $c \ge 2$, it is shown that the Nash equilibrium for the syndicate and crowd occurs when the syndicate chooses $s = t$ and the crowd chooses quick picks from an uniform distribution. It is also shown that small coordinating groups in the crowd have little impact on the syndicate's returns.

Finally, the equiprobable condition is dropped, there is a take on the betting pool and there is a carryover pool. We prove that (1) the syndicate can always achieve a better-than-fair split of the jackpot pool, (2) the best asymptotic strategies for the syndicate and crowd consist of betting aligned with ticket probabilities (probability-proportional betting), (3) the syndicate's expected return is greatest when the lottery is equiprobable, and (4) if a crowd is risk averse or risk seeking, its asymptotic expected return is lower than it would be with a probability-proportional strategy. 

%************************************************************************************************

\section{Background and Related Work}

The fact that the syndicate has a winning strategy is due to the basic logic of coalition formation in games (\cite{myerson1997game}). Each player $i \in N$ in a noncooperative game has a ``reserve'' value $\nu_i$ that consists of the minimum payoff the player can achieve when acting alone. But to each coalition $S$ of players, there will be a total payoff achieveable through cooperation, $\nu(S)$, which is never less than the sum of the reservation values $\nu_i$ of coalition members, $\sum_{i \in S} \nu_i$. We show that a lottery game in which players in $S$ choose to act as a coalition and ones in $N \, \backslash \, S$ choose to act independently produces an excess expected value for the coalition: $\nu(S) > \sum_{i \in S} \nu_i$. There is no surprise here --- it is quite plausible that coalitions will have edges. 

An early contribution to betting strategy in government-sponsored lotteries is \cite{chernoff1980analysis}. As a statistician, Chernoff knew about ``digit preference'' in, for example, reported ages in censuses -- there are too many reported ages ending in '$0$' and '$5$'. Using the Massachusetts State Lottery having numbers 0001 to 9999, Chernoff found that numbers ending in '$0$' and '$9$' were unpopular, and he surmised that tickets combining many unpopular numbers might be underrepresented in lottery betting pools. If this were the case, then betting only greatly underrepresented numbers would constitute a winning betting system. Following up on this idea, \cite{ziemba1986dr} report many other numbers that are unpopular in the Canadian ``6/49'' Lottery using weekly data of marginal number frequencies published by the British Colombia and Western Canada Lottery Corporations. Among the unpopular numbers in the Canadian 6/49 lotto were 1, 10, 20, 28, 29, 30, 32, 34 and all but two numbers over 38, giving a total of 19 statistically unpopular numbers. The authors cite popularity of birthdate months and days and the geometry of the ticket layout for the majority right-handed players as reasons for persistent popularity and unpopularity, and recognize that some numbers can be temporarily overbet when they penetrate the public's awareness. They found that unpopular numbers were stable over time, but in later years, Ziemba's finding that the unpopularities had changed somewhat and had regressed toward the mean provided evidence of a modest learning effect (Ziemba, personal communication). Using empirical probabilities available for the 6/49 numbers and assuming approximate independence  of number probabilities and no carryover pool, \cite{ziemba1986dr} estimate the expected return per dollar wagered for a generic ticket $(i_1, i_2, \ldots, i_6)$ as
\begin{equation}
  \text{Expected Return} = \$0.45 \cdot F_{i_1} \cdot F_{i_2} \cdot F_{i_3} \cdot F_{i_4} \cdot F_{i_5} \cdot F_{i_6},  \label{Eq:ZiembaFactorEquation}
\end{equation}
where the factor $\$0.45$ adjusts for the lottery take and consolation pools, and each factor $F_i$ measures the ratio of equal number probabilities (=$1/49$) to $i$'s estimated betting probability. In lotteries with carryover pools, the factor $\$0.45$ is higher. 

Using formula \eqref{Eq:ZiembaFactorEquation}, the authors investigated the strategy of betting tickets involving only the 19 unpopular numbers and subsets thereof, and find that those strategies win so rarely as to be unattractive as practical systems. \cite{RePEc:inm:ormnsc:v:38:y:1992:i:11:p:1562-1585} investigate an optimal 6/49 Lotto strategy that bets small fractions (to maximize the expected logarithm of final wealth subject to a ruin-avoidance condition) of one's capital using a variant of formula \eqref{Eq:ZiembaFactorEquation}. The authors come to a discouraging conclusion --- that it takes millions of years to achieve a favorable result with high probability!  

We discuss next the literature relevant to our paper's formulas. \cite{10.2307/2117538} show a formula for the expected value of a syndicate
\begin{equation}
  E[\text{syndicate bet of $W$ different tickets}]= \frac{W}{N} \; [R + kc(W + N)](1- e^{pN}) \label{Matheson:LottoFormula}
\end{equation}
where $W$ is the number of tickets bet by the syndicate, $c$ is the cost of one combination, $k$ is the fraction of the handle going into the jackpot, $N$ is number of bets by the crowd, $R$ is the carryover from previous drawings, and $p$ is the probability of winning in a single play. Using the notation of Section \ref{S:Lotteries}, formula \eqref{Matheson:LottoFormula} with $c=1$ is essentially a Poisson approximation to formula \eqref{E:SyndicateExpectedWin-s} multiplied by the jackpot
\begin{equation}
  a + (s + c)(1 - x). \label{JackpotForumla}
\end{equation}
The authors do not mention that expectation \eqref{Matheson:LottoFormula} is positive when expression \eqref{JackpotForumla} $> 0$ and $k=1$, the main assertion of our paper. The purpose of their paper was to advise on economies of scale in lotteries (e.g., multi-state lotteries), not to discuss \eqref{Matheson:LottoFormula} as a potential winning strategy. Of the papers surveyed here, \cite{970220123219961201} provide the most complete account of the expected values for lottery strategies, including the purchase of one of each ticket, a strategy the authors call the ``trump ticket.'' Their formulas are then used to evaluate the ``fairness'' of lotteries in the context of externalities. No analysis of optimal strategies or recognition of the strategic value of a trump ticket is discussed. \cite{RePEc:hcx:wpaper:1109} also discuss the trump ticket. While the authors acknowledge that the return per ticket is generally better when the trump ticket is bet than when a single ticket is bet, there is no discussion of the trump ticket as a potentially winning strategy or its role in a Nash equilibrium.

Much empirical research on lotteries has been done on racetrack parimutuel pools and sports betting. These studies agree that (a) parimutuel odds are consistent with race or game outcomes \cite{citeulike:81468} and (b) there is a persistent favorite-longshot bias (FLB) in individual races which results in underbetting of favorites and overbetting of longshots (\cite{Ziemba2008183}). Several non-mutually exclusive explanations have been offered to explain the FLB bias: (1) poor estimation of probabilities, (2) inside bettors, (3) preference for risk or skewness, (4) heterogeneous beliefs, (5) market power of an uninformed bookmaker, (6) constrained arbitrage and (7) last minute betting by informed players. To this we add a mitigating factor: the presence or absence of syndicates. Many other details of Pick 6's and other lottery-like racetack parlays can be found in \cite{Ziemba:Adventures,Ziemba:ExoticBetting}.

A sizable segment of the literature discusses the possibility of winning betting systems at the racetrack or in sports, such as \cite{RePEc:inm:ormnsc:v:27:y:1981:i:12:p:1435-1452} and \cite{citeulike:1337256}, but they concentrate mostly on individual races or simple parlays, not on full-blown lotteries. 

%************************************************************************************************
\section{Lotteries}\label{S:Lotteries}

In this paper we consider \emph{pure jackpot} lotteries that award only one prize, a jackpot that consists of a \emph{carryover pool} (possibly 0) from the previous drawing and a fraction of the monies wagered for the current drawing. The jackpot is shared equally among all who hold the winning ticket selected at the drawing. If no one holds the winning ticket, the current jackpot pool becomes the carryover pool for the next drawing. 

There are two groups of bettors: a \emph{syndicate} that coordinates its betting and a \emph{crowd} that does not; the manner of (in)coordination is described below. We use the following notation:
\begin{itemize}
\item{$t$,} the number of tickets in the lottery, each costing $\$1$.
\item{$D$,} the random winning ticket drawn using probabilities $P[D = i] = p_i$, $i=1,\ldots,t$.
\item{$s$,} the total number of tickets purchased by the syndicate ($s \ge 1$), with ticket $i$ purchased in amount $s_i = s \cdot r_i$, where $\sum r_i \ge 1$. The syndicate's number of winning tickets is $s_{\!_D}$.
\item{$c$,} the number of individuals in the crowd, each betting one ticket using probabilities $q = (q_1, q_2, \ldots, q_t)'$.
\item{$K$,} a multinomially distributed $t$-vector $K \sim Multin(c,q)$ for the numbers of tickets held by the crowd. $K_i$ is the number of tickets bet on ticket $i$, and $K_{\!_D}$ is the number bet on the winning ticket.
\item{$v$,} the cash jackpot $v = a + (s + c)(1 - x)$,  where $a \ge 0$ is the carryover pool and $x$ is the the (fractional) take on the \emph{betting pool} of $s + c$. 
\end{itemize}

Since the marginal distribution of each component of a multinomial distribution is binomial, the distribution of the number of winning tickets held by the crowd is  
\begin{equation*}
  P[ K_{\!_D} = k ] = 
  \binom{c}{k}
  q_{\!_D}^k \left( 1 - q_{\!_D} \right)^{c-k},
  \quad k = 0, 1, \ldots, c. \label{E:DistributionOfCrowdNbrOfWinners}
\end{equation*}
We define a probability $t$-vector $e_s$ by
\begin{equation}
   e_s = \begin{cases} \label{E:e_s}
           1/s & \text{if } 1 \le i \le s \\
           0   & \text{if } s < t \text{ and } s+1 \le i \le t
         \end{cases}
\end{equation}
so that 
\begin{equation}
  e_t = t^{-1} 1, \label{E:e_t}
\end{equation}
where $1$ is a $t$-vector of ones.

Thus the syndicate will hold $s_{\!_{D}}$ and the crowd, $K_{\!_{D}}$ winning tickets. The random win $W(s,c,p,r,q,a,x)$, gain $G(s,c,p,r,q,a,x)$ and return $R(s,c,p,r,q,a,x)$ to the syndicate are
\begin{align}
  W(s,c,p,r,q,a,x) &= v \frac{s_{\!_{D}}}{s_{\!_{D}} + K_{\!_{D}}}, \label{E:GenericRandomSyndicateWin} \\
  G(s,c,p,r,q,a,x) &= W(s,c,p,r,q,a,x) - s, \label{E:GenericRandomSyndicateGain} \\
  R(s,c,p,r,q,a,x) &= G(s,c,p,r,q,a,x) / s. \label{E:GenericRandomSyndicateReturn}
\end{align}
where it is understood that a ratio $0/(0 + 0) = 0$ in \eqref{E:GenericRandomSyndicateWin}. The syndicate's \emph{expected gain} and \emph{expected return} are
\begin{align}
  E[ G(s,c,p,r,q,a,x) ] &= v \, E \left[  W(s,c,p,r,q,a,x) \right]  - s \nonumber \\
                        &= v \, \left\{ \sum_{i=1}^{i=t} P[D = i] E[ W(s,c,p,r,q,a,x) \, | \, D = i] \right\}  - s \nonumber \\
                        &= v \, \left\{ \sum_{i=1}^{i=t} p_i \sum_{k=0}^{k=t}
                                        \binom{c}{k}
                                      q_i^k \left(1 - q_i \right)^{c-k} \frac{s_i}{s_i \, + \, k} \right\} - s, \label{E:GenericExpectedGain} \\
  E[ R(s,c,p,r,q,a,x) ] &= E[ G(s,c,p,r,q,a,x) ] / s, \nonumber %\label{E:GenericExpectedReturn}
\end{align}
where $v = a + (s + c)(1-x)$ is the jackpot and $s_i/(s_i \, + \, k) = 0$ when $s_i = k = 0$ in \eqref{E:GenericExpectedGain}. In the following, we use notation such as $G(s)$ or $R(s,r)$ to indicate that we study \eqref{E:GenericRandomSyndicateWin}, \eqref{E:GenericRandomSyndicateGain} or \eqref{E:GenericRandomSyndicateReturn} as functions of the variables indicated, other variables having specified values. 

%************************************************************************************************
\section{A Simple $1000$ Number Lottery}\label{SS:WinningAtTheLottery} 

Suppose you are strolling in the park one fine day and see a lottery stand offering a one-day special. The proprietor informs you that she will return in equal shares to the winners all monies bet --- she will reserve no portion for herself. She explains that there are \(1,000\) tickets each costing \(\$1\), and that \(1,000\) people have already bet a \$1 ``quick pick'', a ticket chosen randomly from an uniform distribution with replacement. She says that currently, this $\$1,000$ constitutes the entire prize pool and if no one wins it, it becomes part of tomorrow's jackpot. She says she will close the betting soon and asks if you want to participate. You just happen to have \(\$1,000\) in your pocket. Should you bet?
\mbox{ } \\

We assume that your goal is to maximize your expected return. At first thought, betting seems unwise because the lottery is ``fair'' in the sense that everyone has the same opportunity to win. On the other hand, your composite opponent has played a very bad strategy indeed -- that of picking each lottery ticket randomly, independent of previous choices. That scheme of picking tickets leads by chance to some numbers being bet twice, some three times and some not at all. 

But that scheme is inferior to always picking unbet tickets. To see this, imagine you act as the crowd's proxy and will bet the \(\$1,000\) for them. Suppose \(s < 1,000\) different tickets have been bet and consider how to bet the $(s+1)$-st, either on (a) a ticket already bet (the \(i^{th}\)), or (b) on one not yet bet (the \(j^{th}\)). Which choice yields the greater expected return? Choices (a) and (b) have the same payoffs on every ticket drawn except for the \(i^{th}\) and \(j^{th}\). Assume that the distribution of tickets the crowd bets on the \(i^{th}\) and \(j^{th}\) are the same, and \(k\) are already bet. If the \(i^{th}\) ticket is drawn, the payoffs to the two strategies are (a) \(2/(k+2)\) and (b) \(1/(k+1)\), while if the \(j^{th}\) is drawn, the  payoffs are (a) \(0\) and (b) \(1/(k+1)\).  Because these two possibilities are drawn with the same probability, their sums can be compared --- that of (a)'s two cases, \(2/(k+2)\) with (b)'s, \(2/(k+1)\). But \(2/(k+1) > 2/(k+2) \) showing that it is always better to choose an unbet ticket.

This reduces considerably the candidate strategies. Only those which bet $s$ different tickets, \(0 \le s \le 1,000\), are admissible. Since in an equiprobable lottery, every set of $s$ different tickets occurs with the same probability, the particular tickets selected do not affect calculations of expectations. Using the notation from Section \ref{S:Lotteries} with $a = x = 0$, $t = c = 1000$, $p_i = q_i = 1/t$ and $s_i$ as above, we calculate $R(s)$, the syndicate's return from betting $s$ different tickets.

\begin{enumerate}[label={(\roman*)},ref={\thethm~(\roman*)}]

\item Setting $z = 1/1000$, using the identity
\begin{equation}
  \frac {1} {1 + k}
 \binom{1000}{k}
  =
  \frac {1} {1001}
  \binom{1001}{k+1} \label{E:CombinatorialIdentity}
\end{equation}
and applying formula \eqref{E:GenericExpectedGain} gives expected payoff
\begin{align}
  E[ W(s) ] & = (1000 + s) \left\{ \sum_{i=1}^{i=s} z \sum_{k=0}^{k=1000} \binom{c}{k} z^k (1 - z)^{1000 - k} \frac{1}{1+k} \right\} \label{E:Step1ExpectedWin(s)} \\ 
            & = (1000 + s) s \left\{ 
                     \sum_{k=0}^{k=1000} 
                       \frac {1} {1001}
                       \binom{1001}{k+1}
                       z^{k+1} \left( 1 - z \right)^{1001-(k+1)} \right\} \label{E:Step2ExpectedWin(s)}
                \\
            & = \frac {(1000 + s)s} {1001}
                     \Bigg\{ \left( \sum_{k=0}^{k=1001} 
                     \binom{1001}{k}
                     z^{k} \left( 1 - z \right)^{1001-k} \right) \nonumber \\
            & \phantom{=} \qquad  - (1 - z)^{1001} \Bigg\} \label{E:Step3ExpectedWin(s)} \\
            & = \frac {(1000 + s)s} {1001} \left( 1 - \left( 1 - \frac{1}{1000}  \right)^{1001} \right). \label{E:Step4ExpectedWin(s)}
\end{align}
In \eqref{E:Step1ExpectedWin(s)}, the identical terms in the inner sum allows the factor $s$ to be moved outside the sum and $z$ to be merged with $z^k$ in \eqref{E:Step2ExpectedWin(s)}. By applying the identity \eqref{E:CombinatorialIdentity} in \eqref{E:Step2ExpectedWin(s)} and adding and subtracting the term $(1 - z)^{1001}$ in \eqref{E:Step3ExpectedWin(s)}, the expression \eqref{E:Step4ExpectedWin(s)} is obtained.

\item Using \eqref{E:Step4ExpectedWin(s)}, expresions for $E[G(s)]$ and $E[R(s)]$ are
\begin{align}
  E[G(s)] & = E[W(s)] - s \nonumber \\
          & = \frac {(1000 + s)s} { 1001} \left( 1 - \left(1 - \frac{1}{1000} \right)^{1001} \right) - s. \label{E:ExpectedGain} \\ 
  E[R(s)] &= \frac {1000 + s} { 1001} \left(1 - \left( 1 - \frac{1}{1000} \right)^{1001}\right) - 1. \label{E:ExpectedReturn}
\end{align}

\item Equation \eqref{E:ExpectedGain} is quadratic in $s$ and has a positive second derivative (Figure \ref{G:LotteryGains}, Top Panel.) Thus there is a minimum. Solving the for the minimum gives \(s^* = 290.7981\) with \(E[G(s^*)] = -53.55\). The first \(s\) such that \(E[ G(s) ] > 0\) is \(s_0 = 583\). The return \eqref{E:ExpectedReturn} is linear with respect to the amount bet, has a positive slope, and therefore has maximum of \(26.41\%\) at \(s = 1000\) (Figure \ref{G:LotteryGains}, Bottom Panel.)
\begin{figure}[ht!]
  \centering
    \includegraphics[scale=0.4]{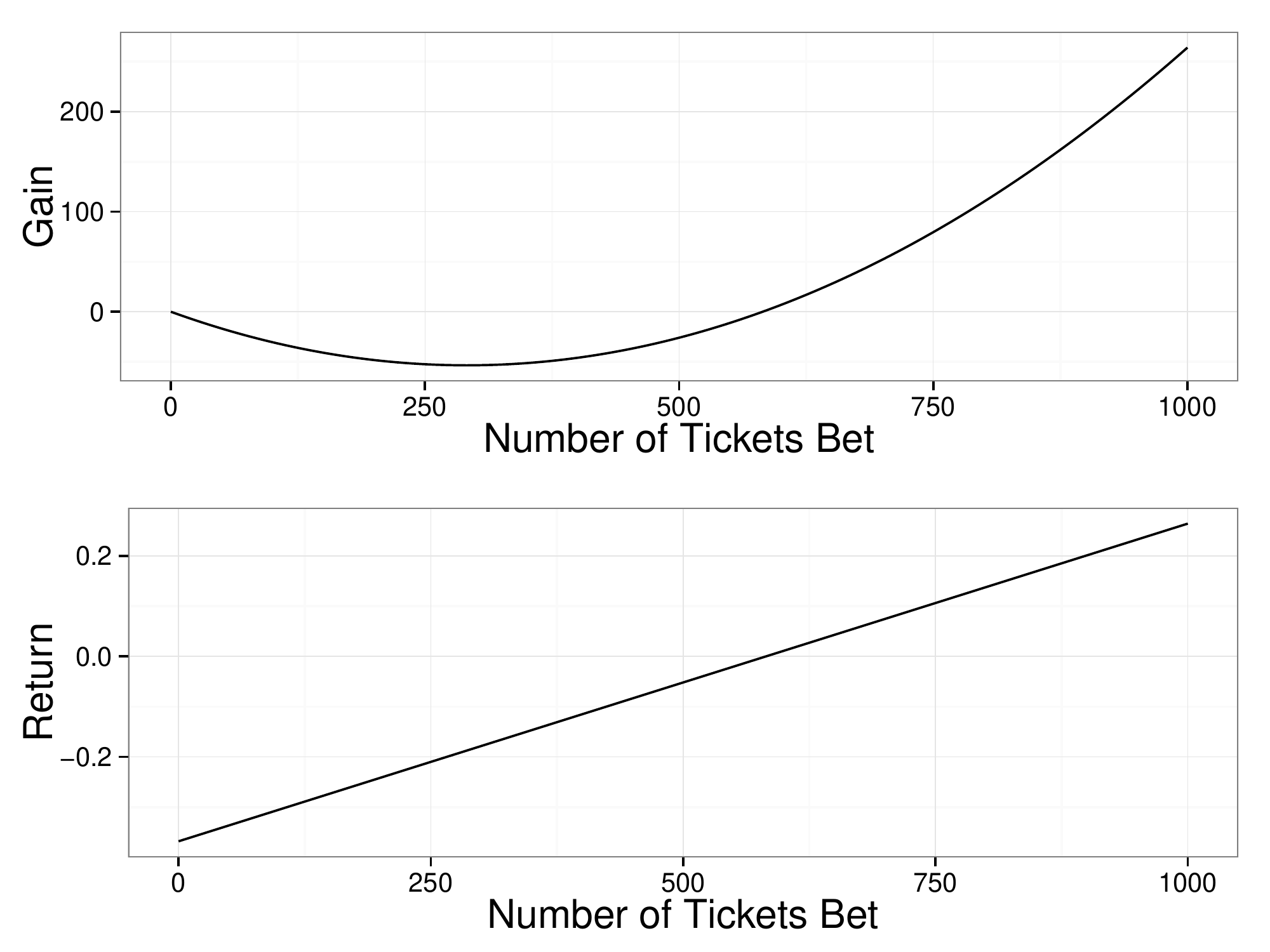}
    \caption{\small{Plots of Gains From Lottery: (A) Top Panel: Net Gains as a 
             function of Number of Tickets Bet using formula 
             \eqref{E:ExpectedGain}, (B) Percentage Return as a Function 
             of Number of Tickets Bet using formula 
             \eqref{E:ExpectedReturn}.}}\label{G:LotteryGains}
\end{figure}
What is remarkable about this result is not so much the edge, but its magnitude, \(26.41\%\)!

\item Figure \ref{G:LotteryGainsAsAFunctionOfCrowdsBettingSize} shows the effect modifying the crowd's betting size using formula \eqref{E:SyndicateExpectedWinrowdProp-t} developed later. 
\begin{figure}[ht!] 
  \centering
    \includegraphics[scale=0.35]{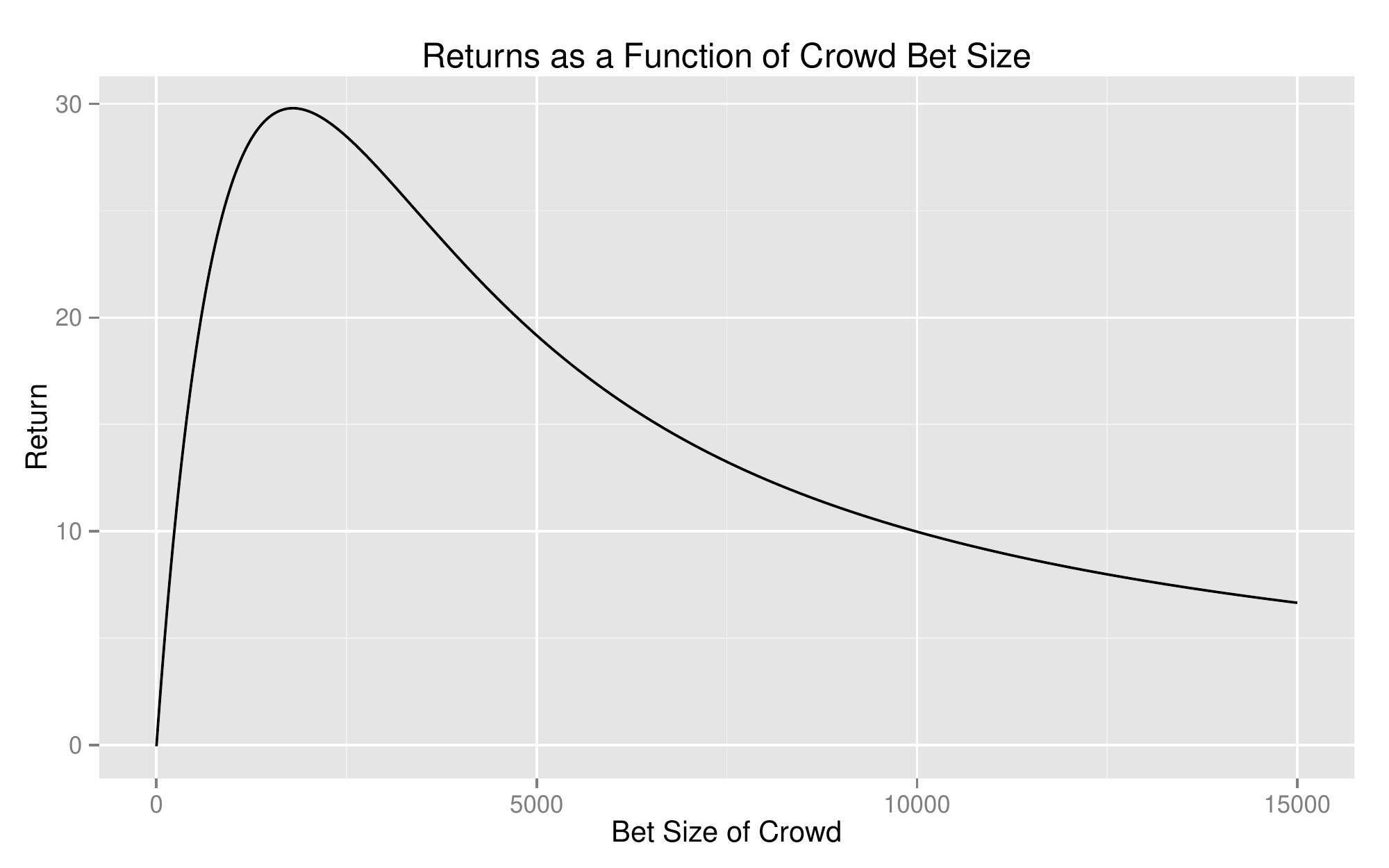}
    \caption{\small{Plots of returns to the syndicate in a 1,000 number lottery as a 
             function of the size of the crowd's bet.}}
             \label{G:LotteryGainsAsAFunctionOfCrowdsBettingSize}
\end{figure}
Note the slow die-off of the return as the crowd increases its betting size. The return is about $19\%$ if the crowd bets $\$5,000$ and is about $10\%$ if the crowd bets $\$10,000$.

\end{enumerate}

%************************************************************************************************
\subsection{Intuition Behind the Syndicate's Win}\label{SS:IntuitionBehindTheFTCEffect}

Why does the crowd lose? The answer: as a composite player, theirs is an inferior strategy. The only ways they can avoid this problem are to subscribe to a coordination mechanism that doesn't duplicate the tickets they bet (cooperate) or to bet nothing (the Nash equilibrium when utility is linear). Such a mechanism could be provided, for example, by a quick-pick machine that selects a random number without replacement, i.e., that explicitly avoids giving duplicates. But of course, the decision to use such a machine is a cooperative act which is merely facilitated by a machine.

Second, why is there no profit until at least \(\$583\) has been bet? The reason is subtle, but informative. Consider for a random ticket, the probability distribution of tickets held by the crowd.  These are
\begin{equation*}
  P[K=k] = \binom{1000}{k}
           \frac{1}{1000}^k \left( 1 - \frac{1}{1000} \right)^{1000-k}. \label{E:PoissonApproximation1000} \\
\end{equation*}
Table \ref{Ta:LotteryExpectedGain} shows the probabilities and the syndicate's expected winnings as a function of the crowd holding $4$ or fewer winning tickets. The first column shows the number of winning tickets ($k$) held by the crowd, the second, the probability $P[K=k]$ that the crowd holds $k$ winning tickets, the third and fourth, the amount won and the $k$-contribution to the expected payoff ({$E[W_{1000}]$) for a syndicate that bets $s=1000$ different tickets, respectively, and the fifth and sixth, the amount won and the $k$-contribution to the expected payoff ({$E[W_{1}]$) for a syndicate that bets $s=1$ ticket, respectively.
\begin{table}[ht]
  \begin{center}
    \caption{\small{Expected Gain as a Function of Tickets Held by the Crowd.}}
      \label{Ta:LotteryExpectedGain}
    \small
    \begin{tabular} {cccccc}
      \rule[-4pt]{0pt}{10pt} \\
      \hline
      \rule[-4pt]{0pt}{10pt} \\
      \multicolumn{1}{c}{(1)} &  \multicolumn{1}{c}{(2)}      &  \multicolumn{1}{c}{(3)}                  & \multicolumn{1}{c}{(4)}                       &  \multicolumn{1}{c}{(5)}            & \multicolumn{1}{c}{(6)} \B \\
      \multicolumn{1}{c}{}    &  \multicolumn{1}{c}{}         &  \multicolumn{1}{c}{Syndicate}            & \multicolumn{1}{c}{$E[W_{1000} \; | \; K=k]$} &  \multicolumn{1}{c}{Syndicate}      & \multicolumn{1}{c}{$E[W_{1} \; | \; K=k]$} \\
      \multicolumn{1}{c}{$k$} &  \multicolumn{1}{c}{$P[K=k]$} &  \multicolumn{1}{c}{Payoff ($s=1000$)}    & \multicolumn{1}{c}{$ * P[\, K=k]$}            &  \multicolumn{1}{c}{Payoff ($s=1$)} & \multicolumn{1}{c}{$ * P[\, K=k]$}     \\
      \rule{0pt}{2pt} \\
      \hline
      \rule{0pt}{0pt} \\
      $0$ & $0.368$ &        $\$2000$  & \phantom{1}\$735.76 &           \$1001.0 & $\$0.368$ \\
      $1$ & $0.368$ &        $\$1000$  & \phantom{1}\$367.88 & \phantom{1}\$500.5 & $\$0.184$ \\
      $2$ & $0.184$ & \text{ $\$666$}  & \phantom{1}\$122.63 & \phantom{1}\$333.7 & $\$0.061$ \\
      $3$ & $0.061$ & \text{ $\$500$}  & \phantom{11}\$30.66 & \phantom{1}\$250.3 & $\$0.015$ \\
      $4$ & $0.015$ & \text{ $\$400$}  & \phantom{111}\$6.13 & \phantom{1}\$200.2 & $\$0.003$ \\
      \rule{0pt}{0pt} \\
      \hline
      \rule{0pt}{0pt} \\
          &         &  Sum             &           \$1263.05 &                    & $\$0.632$ \\
    \end{tabular}
  \end{center}
\end{table}
A syndicate that bets only $\$1$ has expected payoffs of only $\$0.368$, $\$0.184$, etc. for a total of $\$0.632$ as shown in column (6). This is short of the syndicate's \$1 bet by $\$1 -  \$0.632 = \$0.368$ which is the expected amount that goes into the carryover pool (and approximately, the probability of $k=0$) when $1001$ tickets are bet at random. But a syndicate that bets $s = 1,000$ reduces the probability of a carryover to zero, so that the term $k=0$ contributes $0.73576$ per dollar bet compared to $0.368$ per dollar bet when $s=1$. Not until the syndicate bets $291$ are there enough tickets unbet by the crowd, that the marginal return on the next ticket is greater than zero. After $291$, the probability of betting a ticket the crowd does not hold increases with each additional $\$1$ and the expected return therefore increases monotonically beyond that number due to succesive reduction of the probability that there are no winners.

%************************************************************************************************
\section{Equiprobable Lotteries without Carryover or Take}\label{S:EquiprobableLotteriesWithoutCarryoverOrTake} 

\subsection{The Syndicate Bets $\bm{s \le t}$}

Using the notation of Section \ref{S:Lotteries}, we study lotteries for which $v = s + c$, $p = t^{-1} 1$, $a = x = 0$,  and $r = e_s$, where $e_s$ is defined in \eqref{E:e_s} and \eqref{E:e_t}. Our main interest is the returns from various choices of $s$, $c$, $q$ and $r$:
\[
  R(s,c,r,q) = R(s,c,p=t^{-1} 1,r,q,a=0,x=0)
\]
These lotteries have most of the important characteristics of more complicated ones, but are easier to analyze. In the initial part, we assume that each group in the crowd consists of a single bettor. This condition is relaxed in Proposition \ref{Prop:SmallCoordinatingGroups}.

A lemma is useful for the main theorem. Its straightforward defivation follows that of \eqref{E:Step1ExpectedWin(s)}-\eqref{E:Step4ExpectedWin(s)} and is omitted.
\begin{lem}[Expected Value for \textit{W(s,c,e\textsubscript{s},q)}]\label{Lem:ExpectedW(t,c,q,1/t1)} For $s \le t$ and $e_s$ is defined in \eqref{E:e_s},
\begin{equation*}
  E[ W(s,c,e_s,q) ] = \frac{c+s}{c+1} \frac{s}{t} \left( \frac{1}{t} 
                           \sum_{i=1}^{i=t} 
                           \frac{1}{q_i} \left( 1 - \left( 1 - q_i \right)^{c + 1} \right) \right). \label{E:SyndicateExpectedWin-s}
\end{equation*}
\end{lem}
\noindent The main result is 
\begin{thm}[Optimal Syndicate and Crowd Betting] \label{Thm:OptimalSyndicateAndCrowdBetting}
With $e_t = t^{-1} 1$ and with assumptions $p = t^{-1} 1$, $a = x = 0$, $c \ge 2$, and $s \le t$, and using the notation $R(s,c,r,q)$ for the syndicate's return, 
\begin{align}
  R(s,c,e_s,q)   \ge R(s,c,e_s,e_t) \quad &\text{ for } 1 \le s \le t, \text{ with equality if and only if } q = e_t. \label{E:CrowdOptimum} \\
  R(t,c,e_t,e_t) \ge R(s,c,e_s,e_t) \quad &\text{ for } 1 \le s \le t, \text{ with equality if and only if } s = t. \label{E:SyndicateOptimum} \\
  R(t,c,e_t,q) &> 0 \quad \text{ for any } q. \label{E:SyndicateEdge}
\end{align}
\end{thm}

\noindent \textit{Proof:}  From Lemma \ref{Lem:ExpectedW(t,c,q,1/t1)}, we determine that
\begin{align}
  E[ R(s,c,e_s,q) ] &=  E[ W(s,c,e_s,q) ] / s   - 1 \nonumber \\
                    &= \frac{c + s}{c + 1} \frac{1}{t^2} 
                           \left( \sum_{i=1}^{i=t} 
                           \frac{1}{q_i} \left( 1 - \left( 1 - q_i \right)^{c + 1} \right) \right) - 1. \label{E:E[R(s,c,e_t,q)]} 
\end{align}
The functions
\begin{equation}
  f(q) = q^{-1} \left( 1 - \left( 1 - q \right)^{c+1} \right) \label{E:ExpectedReturnFor1/1+k}
\end{equation}
in equation \eqref{E:E[R(s,c,e_t,q)]} are positive, strictly decreasing and convex on $(0,1)$ for $c \ge 2$ (Appendix \ref{S:AnalysisOfEquation1/1+X}). Since a sum of $t$ (strictly) convex functions on $(0,1)$ is a (strictly) convex function on $(0,1)^t$ and remains (strictly) convex on any convex subset of $(0,1)^t$, it follows that the sum \eqref{E:E[R(s,c,e_t,q)]} is strictly convex on the simplex
\[
  \left\{ \: q \in (0,1)^t \; \bigg| \; q \ge 0, \, \sum_{i=1}^{i=t} q_i = 1 \right\}.
\]
Further, a constrained optimization of the sum in \eqref{E:E[R(s,c,e_t,q)]} yields the first order conditions
\[
  - \frac{1 - (1 - q_i)^{c + 1}}{q_i^2} + \frac{(c + 1)(1 - q_i)^{c}}{q_i} - \gamma = 0,
\]
for constant $\gamma$ and $i = 1, 2, \ldots, t$. These conditions are satisfied for $q_i = 1/t$, which in view of previous remarks shows this to be the unique minimum of \eqref{E:E[R(s,c,e_t,q)]} in the simplex. We have thus shown \eqref{E:CrowdOptimum}:
\begin{align}
  E[ R(s,c,e_s,q)] &= \frac {c+s}{c+1} \frac{1}{t^2} \left(  
                        \sum_{i=1}^{i=t} 
                        \frac{1}{q_i} \left( 1 - \left( 1 - q_i \right)^{c + 1} \right) \right) - 1 \nonumber \\
                   &\ge \frac{(c+s)}{c+1} \left( 1 - \left( 1 - \frac{1}{t} \right)^{c + 1} \right) - 1 \nonumber \\
                   &= E[ R(s,c,e_s,e_t)] \label{E:OptimalityOfCrowdBettingForslet}
\end{align}
with strict inequality when $q \ne e_t$, which proves inequality \eqref{E:CrowdOptimum}. 

For inequality \eqref{E:SyndicateOptimum}, $E[ R(s,c,e_s,q) ]$ as a function of $s$ on $[1,t]$ is linear with positive slope. Therefore, it achieves its unique maximum at $s = t$.

Since the syndicate's edge is minimized when $q = e_t$ (by \eqref{E:CrowdOptimum}) and its gain maximized when $s = t$ (by inequality \eqref{E:SyndicateOptimum}), inequality \eqref{E:SyndicateEdge} will be demonstrated if 
\begin{equation}
 E[ R(t,c,e_t,e_t) ] = \frac{c + t}{c + 1}  
                       \left( 1 - \left( 1 - t^{-1} \right)^{c + 1} \right) - 1 > 0 \label{E:SyndicateExpectedWinrowdProp-t}
\end{equation}
is true. But rearranging \eqref{E:SyndicateExpectedWinrowdProp-t},
\begin{align}
  E[ R(t,c,e_t,e_t) ] > 0 &\iff \left( 1 - \left( 1 - t^{-1} \right)^{c + 1} \right) > \frac{c + 1}{t + c} \nonumber \\
                          &\iff \left( 1 - \frac{1}{t} \right) ^ {c + 1} < \quad \frac{t - 1}{t + c} \nonumber \\
                          &\iff c \cdot log \left( \frac{t-1}{t} \right) + log \left(\frac{t + c}{t}\right) < 0. \label{E:ReturnInequalityForSyndicateEdge}
\end{align}
To verify inequality \eqref{E:ReturnInequalityForSyndicateEdge}, let $g(x)$ be the strictly concave function
\[
 g(x) = log \left( \frac{t + x}{t} \right)
\]
and $X$ the random variable
\begin{equation*}
  X = \begin{cases}
        -1 & \text{ with probability } \, \frac{c}{c+1}, \\
         c & \text{ with probability } \, \frac{1}{c+1}.
      \end{cases}
\end{equation*}
Then using Jensen's inequality,
\begin{equation}
   \frac{c}{c+1} log \left( \frac{t - 1}{t} \right) + \frac{1}{c+1} log \left( \frac{t + c}{t} \right) = 
       E[g(X)] < g(E[X]) = g(0) = 0. \label{E:Jensens'EqnForSyndicateEdge}
\end{equation}
Multiplying both sides of \eqref{E:Jensens'EqnForSyndicateEdge} by $c+1$ then gives \eqref{E:ReturnInequalityForSyndicateEdge}.

$\blacksquare$

\begin{cor}[Nash Equilibrium of the Lottery Game] With the same setup as in Theorem \ref{Thm:OptimalSyndicateAndCrowdBetting}, the strategy $s = t$ for syndicate and $q = e_t$ for crowd is the unique Nash equilibrium:
\begin{equation*}
  R(t,c,e_t,q) \ge R(t,c,e_t,e_t) \ge R(t,c,r,e_t), \label{E:SyndicateNash}
\end{equation*}
and the syndicate has a positive expected return at equilibrium.
\end{cor}

\noindent \textit{Proof:} Immediate from Theorem \ref{Thm:OptimalSyndicateAndCrowdBetting}.

$\blacksquare$

We address some further questions about the syndicate's expected return in the next two propositions.
\begin{prop}[Breakeven Bet Sizes] \label{Prop:BreakevenBetSizes} Using the same set up as Theorem \ref{Thm:OptimalSyndicateAndCrowdBetting} and assuming $q = e_t$,

\begin{enumerate}
\item The syndicate's minimum expected gain occurs at $s^* = \frac{1}{2} \left( \frac{1 + c y}{1 - y} \right)$. \label{E:SyndicateMinimums} 
\item The syndicate's breakeven expected gain occurs at $s_0 = 2s^*$.
\end{enumerate}

\end{prop}

\noindent \textit{Proof:} The exact expected gain from \eqref{E:OptimalityOfCrowdBettingForslet} is
\[
  g(s) = E[ G(s,c,e_s,e_t)] = \frac{(c+s)s}{c+1} \left( 1 - \left( 1 - \frac{1}{t} \right)^{c + 1} \right) - s
\]
for $s \in [0,1]$ when the crowd bets optimally using $q = e_t$. Setting $y = (1 - \frac{1}{t})^{c+1}$, we calculate the first derivative
\begin{equation*}
  g'(s) = \frac{2 s + c}{1 + c} (1 - y) - 1,
\end{equation*}
and find a single critical point
\begin{equation}
  s^* = \frac{1}{2} \left( \frac{1 + c y}{1 - y} \right). \label{E:SyndicateMinimums}
\end{equation}
Substituting $s^*$ into $g$ gives the minimum value  
\begin{equation}
  g(s^*) = - \frac{1}{4} \frac{(1 + cy)^2}{(1 - y)(1+c)}. \nonumber %\label{E:SyndicateWorstLoss}
\end{equation}
which is $< 0$. Since
\[
  g''(s) = \frac{2}{1 + c} (1 - y) > 0,
\]
$s^*$ is the unique minimum of $g(s)$ and $g$ is convex. The question is: does $s^*$ lie between $0$ and $t$? Since \eqref{E:SyndicateMinimums} is clearly positive, we check for $s^* < t$. But this is clearly the case, since $g(0) = 0$, $g(t) > 0$, $g(s^*) < 0$ and $g$ is strictly convex on $[0,\infty)$. 

Since $g(s)$ is strictly increasing on $[s^*,t]$, $g(s^*) < 0$ and $g(t) > 0$, there is a point at which the gain breaks even. Since $g$ is quadratic in $s$ and symmetric around $s^*$, it follows that $s_0 = 2s^*$ is the break-even point:
\begin{equation*}
  s_0 = 2 s^{*} = \frac{1 + c y}{1 - y}. \label{E:SyndicateFirstPositive}
\end{equation*}

$\blacksquare$

Proposition \eqref{Prop:BreakevenBetSizes} shows that a syndicate's bet of an amount $s \le t$ will have decreasing expected gain for $0 \le s \le s^* = (1 + c y)/(2(1 - y))$, increasing expected gain for $s^* \le s \le t$ and will be positive only if $\floor{s_0} + 1 \le s \le t$.\footnote{ $\floor{y}$ is the floor function, the greatest integer not greater than $y$.}

Betting groups at lotteries are either single individuals who bet a block of different tickets or collections of bettors who pool their funds and act as an individual by selecting different tickets. In either case, a group bets \emph{different tickets}. In order to simplify the analysis, we assume that there are $g$ groups that bet the same number $l$ of different tickets, so that $c = g l$. Assuming that the crowd aligns their betting to the lottery probabilities, the $i^{th}$ ticket will be selected by a group with probability $l/t$ and therefore the number of tickets bet on the $i^{th}$, $K_i$, is distributed binomially: $K_i \sim Bin( g, l/t )$. 
\begin{prop}[Small Coordinating Groups in the Crowd] \label{Prop:SmallCoordinatingGroups} Modify the set up of Theorem \ref{Thm:OptimalSyndicateAndCrowdBetting} by introducing $g$ groups within the crowd that each select $l$ different tickets. Then provided $l \ll min(c,t)$, the syndicate's return is only slightly lower than in the no-group case $g = c$, $l = 1$. 
\end{prop}

\noindent \textit{Proof:} It is shown in Appendix \ref{App:Eq1-1+X}, formula \eqref{E:ApproxE[(1+X)-1]} that
\begin{align*}
  E \left[ \frac{1}{1 + X} \right] \approx \frac{1}{(c+1) q} \left( 1 - exp(-(c+1) q) \right)
\end{align*}
when $c \gg 1$ and $X \sim Bin(c,q)$. Thus
\begin{align}
  E_g[W(s,c,e_t,e_t)] &= (c + s) \left\{ \sum_{i=1}^{i=s} \frac{1}{t} \sum_{k=0}^{k=g} 
                                 \binom{g}{k}
                                 \left( \frac{l}{t} \right)^k \left( 1 - \frac{l}{t} \right)^{g-k} \right\} \nonumber \\
                      & \approx \frac{(c + s)s}{t} \frac{t}{(g+1)l} \left( 1 - exp \left(- \frac{(g+1)l}{t} \right) \right) \nonumber \\
                      & =  \frac{(c + s)s}{c + l} \left( 1 - exp \left(- \frac{c + l}{t} \right) \right). \nonumber %\label{E:ImpactOfGroupBettingOfTheSyndicate'sGain}
\end{align}
When $l = 1$ (so that $g = c$) we get the usual calculation of expectation, and the ratio is
\begin{align}
  \frac{E_g[\, W(s,c,e_t,e_t) \, ]}{E_c[\, W(s,c,e_t,e_t) \, ]} &\approx 
    \frac{\frac{(c + s)s)}{c + l} \left( 1 - exp \left(- \frac{c + l}{t} \right) \right)}
   {\frac{(c + s)s}{c + 1} \left( 1 - exp \left(- \frac{c + 1}{t} \right) \right) } \nonumber \\
             &= \frac{c+1}{c+l} \frac{1 - exp \left(- \frac{c}{t} \right) exp(-l/t) }{1 - exp \left(- \frac{c}{t} \right) exp(-1/t)} \label{E:Ratio}
\end{align}
But under the assumption $l \ll min(c,t)$, \eqref{E:Ratio} will be very close to 1 since 
\begin{align*}
  (c+1)/(c+l) &= (1 + l/c)/(1 + 1/c) \approx 1 \\
  exp(-1/t)   & \approx 1 \\
  exp(-l/t)   & \approx 1.
\end{align*}

$\blacksquare$

Of course, this approximation breaks down if some of the groups have sizes comparable to the syndicate; it would seem that the syndicate's gain is more tied to $\underset{1 \le j \le g}{max} \, c_j$ than an average of the $c_j$, but this idea is not investigated here.

%************************************************************************************************

\subsection{The Syndicate Bets $\bm{s \ge t}$}\label{SS:SyndicateBetss>=t}

\begin{thm}[The Syndicate Bets \textit{$s \ge t$} in an Equiprobable Lottery] \label{Thm:SyndicateBetss>=t} We assume as before that $p = t^{-1} 1$, $a = x = 0$, $c \ge 2$, define $e_t = t^{-1} 1$ and require only that $s \ge t$. We use the notation $R(s,c,r,q)$ for the syndicate's return. Then 
\begin{enumerate}[label={(\roman*)},ref={(\roman*)}]
\item{For any $s \ge t$ and $r$, the crowd minimizes the syndicate's return $R(s,c,r,q)$ by choosing $q = e_t$.} \label{Enum:q=e_t}
\item{If $s = nt$, $n \ge 1$ and $c \ge 2$, the syndicate's optimal strategy bets $n$ on each ticket. In that case $R(nt,c,e_t,q) > 0$, irrespective of $q$}. \label{Enum:s=nt}
\item{For $n \ge 1$, $s > nt$, $s < (n+1)t$ and $c \ge 2$, the syndicate's optimal strategy wagers $n$ on each ticket and $s - nt$ on $s - nt$ different tickets. In this case,  $R(nt,c,e_t,q) > 0$, irrespective of $q$}. \label{Enum:snent}
%\item{Poisson approximations are developed and are easier to work with.}
\end{enumerate}
\end{thm}
The proof is in Appendix \ref{App:BettingStrategiesForSyndicateAndCrowd}.

There is an important message in part \ref{Enum:s=nt} of Theorem \ref{Thm:SyndicateBetss>=t} --- that if $n > 1$ syndicates all bet each ticket once in a lottery with no take and no carryover pool, then each syndicate's expected return is (still) positive! 

%************************************************************************************************
\section{General Non-Equiprobable Lotteries}\label{S:WinningAtNon-UniformLotteries}

In this section, it is assumed that true ticket probabilities have a distribution $P[ D = i] = p_i$, where in general $p \ne e_t$, that the crowd bets using probability vector $q$, the syndicate bets using $r$, and $a, x > 0$. In this section, $s r$ will generally be a fractional vector, since that allowance  produces tractable solutions. In practical usage, though, these fractional solutions must be converted into integral ones, and then examined to ensure that they retain near-optimal properties. 

All results of this section apply without modification to equiprobable lotteries. The winners of a lottery will share the jackpot pool of $v = a + (1-x)(s + c)$, where $a \ge 0$ is a carryover pool, $x$ is the take on the betting pool, $s$ is the amount of the syndicate's bet and $c$ the amount of the crowd's. As in Section \ref{S:Lotteries}, let each crowd member select one ticket independently of everyone else resulting in a random selection $K = (K_1, K_2, \ldots, K_t)'$, $\sum_{i=1}^{i=t} K_i = c$, where $K_i$ the total number bet on the $i^{th}$ ticket. Then the random variable $K$ has a multiomial $Multi(c,q)$ distribution
\[
  P[K_1 = k_1, K_2 = k_2, \ldots, K_t = k_t] = \frac{c!}{k_1! \, k_2! \, \ldots, k_t!} q_1^{k_1} \, q_2^{k_2}, \ldots, q_t^{k_t}.
\]
and the marginal distribution $K_i$ of $K$ is binomial with probability $q_i$: $K_i \sim Bin(c,q_i)$
\begin{equation*}
  P[ K_i = k_i ] = 
    \binom{c}{k_i}
    q_i^{k_i} \left( 1 - q_i \right)^{c - k_i}
\end{equation*}

%************************************************************************************************
\subsection{Lotteries as Games between Syndicate and Crowd}

The contest between syndicate and crowd can be considered as a game in which the probability distribution for tickets is $p$ is known to everyone, the syndicate bets according to $s = s \cdot r$, and the crowd independently according to  $K \sim Multin(c,q)$, where $Multin$ is a multinomial distribution. The game commences with syndicate and crowd selecting tickets, with members of the crowd independently selecting 1 ticket apiece. After ticket selection, a random winning ticket is drawn according to $p$. The payoffs of this game for the syndicate and crowd are their respective expected returns from strategic choices $(r,q)$.

A Nash equilibrium in this game consists of strategies $(r,q)$ for syndicate and crowd such that given $r$, $q$ is a best returning strategy for the crowd and given $q$, $r$ is a best returning strategy for the syndicate. We show below that when $c,s \rightarrow \infty$ and $c/s \rightarrow u$, $u$ a constant, it follows that $r = p$ and $q = p$ asymptotically. Our definition of \emph{risk aversion} and \emph{risk seeking} differ from that standard in game theory and economics and is motivated by the game we have defined, in which the crowd as a whole is considered risk averse or risk seeking since in effect, we treat the crowd as a single stochastic bettor.

\begin{thm}[Results for Non-equiprobable Lotteries] We assume that the syndicate bets a total of $s$ tickets, the crowd bets $c \ge 2$ tickets using probability $t$-vector $q$, that $a, x \ge 0$ and that the syndicate bets $s r_i$ on the $i^{th}$ ticket, where fractional bets are admissible. Then  
\begin{enumerate}[label={(\roman*)},ref={(\roman*)}]
\item{A winning syndicate strategy exists if $a/(s+c) - x \ge 0$,} \label{Enum:NEWinning}
\item{The asymptotic Nash Equilibrium for the syndicate consists of betting with $r = q = p$,} \label{Enum:NEPWinning}
\item{An equiprobable lottery is best for the syndicate, irrespective of the crowd's strategy,} \label{Enum:NEPEquiprobable}
\item{If the crowd is risk averse or risk seeking, its asymptotic return is worse than that of a proportional strategy.} \label{Enum:NEPRisk}
\end{enumerate}
\label{ThmForNon-equiprobableLotteries}
\end{thm}

\noindent \textit{Proof:} Part \ref{Enum:NEWinning}. Suppose that fractional tickets can be bought, and that the syndicate purchases $s$ tickets in fractional amounts $s_i = s p_i$ and let $D$ be the winning lottery ticket. Since $D$ has distribution $P[D = i] = p_i$, the fractional number of tickets bet by the syndicate is $s_{\!_D} = s \, p_{\!_D}$ and the random variable for the crowd's bet is $K_{\!_D}$ with distribution $K_{\!_D} \sim Bin(c , q_{\!_D})$. With 
\begin{align}
  G(s,c,p,r,q) & = \left( a + (1 - x)(s + c) \right) \, \frac{s_{\!_D}}{s_{\!_D} + X_{\!_D}} - s \nonumber \\
               & = v \, \frac{s_{\!_D}}{s_{\!_D} + X_{\!_D}} - s, \nonumber 
\end{align}
the syndicate's expected gain is
\begin{align}
  E[ G(s,c,p,r,q) ] & = v \left\{ \sum_{i=1}^{i=t} P[ \, D = i \, ] \, E\left[ \, \frac{s p_{\!_D}}{s p_{\!_D} + X_{\!_D}} \, \Biggl\vert \, D = i \, \right] \right\} - s \nonumber \\ 
                & = v \left\{ \sum_{i=1}^{i=t} p_i \sum_{k=0}^{k=c} 
                      \binom{c}{k}
                            q_i^k (1 - q_i)^{c - k} \frac{s p_i}{ s p_i + k} \right\} - s \nonumber \\
                & > v \left\{ \sum_{i=1}^{i=t} p_i \frac{s p_i}{s p_i + c q_i} \right\} - s. \label{E:InteriorExpectationInequality1}
\end{align}
where the last step follows from Jensen's inequality. It can be shown using constrained optimization that expression \eqref{E:InteriorExpectationInequality1} is minimized with respect to $q$ at $q = p$. Therefore
\begin{align}
  E[ G(s,c,p,r,q) ] & > v \left\{ \sum_{i=1}^{i=t} p_i \frac{s p_i}{s p_i + c p_i} \right\} - s \nonumber \\
                & = \left( a + (1 - x)(s + c) \right)  \frac{s}{s + c} - s \nonumber \\ 
                & = s \left( \frac{a}{s + c} \, - \, x \right). \nonumber 
\end{align}

\noindent Part \ref{Enum:NEPWinning}. Consider first a lottery with no take in which the syndicate bets proportionally to $p$ and the crowd uses probabilities $q$. What is the best asympotic choice of $q$ to minimize syndicate's expected gain. We calculate $E[G(s,c,p,r,q)]$ as
\begin{align}
  E[ G(s,c,p,r,q) ] & = v \, \left\{ \sum_{i=1}^{i=t} P[ \, D = i \, ] \, E\left[ \, \frac{s_{\!_D}}{s_{\!_D} + X_{\!_D}} \, \Biggl\vert \, D = i \, \right] \right\} - s \nonumber \\
         & = v \, \left\{ \sum_{i=1}^{i=t} p_i \sum_{k=0}^{k=c} 
                   \binom{c}{k}
                   q_i^k (1 - q_i)^{c - k} \frac{s p_i}{ s p_i + k} \right\} - s \nonumber %\label{GeneralSyndicateGainOn-p-q}
\end{align}
The first order conditions require that an optimum $q_i^*$ satisfy 
\begin{equation}
  \beta = c p_i \, \sum_{k=0}^{k=c - 1} 
                      \binom{c-1}{k}
                      {q_i^*}^k (1 - {q_i^*})^{c - 1 - k} (g(i,k+1) - g(i,k)),
            \label{E:LagrangeEquationForqi1}
\end{equation}
for some constant $\beta$ for all $i = 1, 2, \ldots, t$, where 
\begin{equation*}
   g(i,k) = \frac{s p_i}{s p_i + k}.
\end{equation*}
If the lottery is equiprobable ($p_i = 1/t$), then $q_i^* = p_i = t^{-1}$ for all $i$. If the lottery is not equiprobable, then for at least two tickets $i$ and $j$, $p_i \ne p_j$ and from \eqref{E:LagrangeEquationForqi1}, $q_i^* \ne q_j^*$ and the equiprobable argument does not work. In this case, the best choice for $q^*$ will in general not be $p$.  

But if $c, s \rightarrow \infty$ and $c/s \rightarrow u$, $u$ a constant, then 
\begin{align}
  E[ G(s,c,p,r,q) ] & \approx \lim_{\substack{s,c \rightarrow \infty \\
                                         c/s \rightarrow u}} \, \, E[ G(s,c,p,r,q) ] \nonumber \\
                    & = \left( v \sum_{i=1}^{i=t} p_i \frac{1}{1 + u \frac{q_i}{p_i}} \right) \, - s \label{E:LargeSExpectedGain1}
\end{align}
and a Lagrange optimization of $q$ in \eqref{E:LargeSExpectedGain1} yields the equations
\begin{align*}
   \beta & = p_i \left( 1 + u \frac{q_i}{p_i} \right)^{-2} \frac{u}{p_i} \nonumber \\
         & = \left( 1 + u \frac{q_i}{p_i} \right)^{-2} u \label{E:LLN-LagrangeEquations1}
\end{align*}
for a constant $\beta$ and for each $i = 1, 2, \ldots, t$. These equations can be constant only if $q_i / p_i$ is constant for each $i$ which implies $q_i = p_i$.\footnote{Setting \eqref{E:LLN-LagrangeEquations1} equal for $i \ne j$ yields the unique solution $q = p$.}  The Hessian from \eqref{E:LargeSExpectedGain1} has positive entries on the diagonal and zeroes off the diagonal, therefore is positive definite; thus  $q = p$ uniquely minimizes the syndicate's gain. Conclusion: Choosing $q_i = p_i$ for each $i$ minimizes the syndicate's asymptotic expected gain.

Now suppose that the crowd bets using $p$ and the syndicate bets proportionally to $r$. Then 
\begin{align}
  E[ G(r,q=p) ] = v \, \left\{ \sum_{i=1}^{i=t} p_i \sum_{k=0}^{k=c} 
                            \binom{c}{k}
                            p_i^k (1 - p_i)^{c - k} \frac{s r_i}{ s r_i + k} \right\} - s
\end{align}
Then
\begin{align}
  E[ G(r,q=p) ] & \approx \lim_{\substack{s,c \rightarrow \infty \\
                                         c/s \rightarrow u}} \, \, E[ G(r,q=p) ] \nonumber \\
              & = \left( v \sum_{i=1}^{i=t} p_i \left( 1 +  u \frac{p_i}{r_i} \right)^{-1} \right) \, - s \label{E:LargestExpectedGain-r}
\end{align}
and the first order conditions from equation \eqref{E:LargestExpectedGain-r} can be written
\begin{align*}
   \beta & = \left( 1 + u \frac{p_i}{r_i} \right)^{-2} \left( u \frac{p_i}{r_i} \right)^2 \nonumber \\
         & = \left( 1 + \frac{1}{u} \frac{r_i}{p_i} \right)^{-2} \label{E:LLN-LagrangeEquations2}
\end{align*}
and using the same argument as earlier, it follows that $r$ = $p$. Taken together, these results show that, asymptotically at least, the unique best reply of the crowd to $r = p$ is $q = p$ and the unique best reply of the syndicate to $q = p$ is $r = p$, which is precisely a unique Nash equilibrium.

\noindent Part \ref{Enum:NEPEquiprobable}. The general syndicate expected gain is 
\begin{align*}
   G(s,c,p,r,q) &= E_{p,q} \left[ \frac{s p_{\!_D}}{s p_{\!_D} + X_{\!_D}} \right] \\
                &= v \, \left\{ \sum_{i=1}^{i=t} p_i \sum_{k=0}^{k=c} 
                                  \binom{c}{k}
                            q_i^k (1 - q_i)^{c - k} \frac{s p_i}{ s p_i + k} \right\} - s 
\end{align*}
The problem is: What distribution $p$ maximizes $ G(s,c,p,r,q)$ given $q$, $s$ and $c$. In this case no calculation is necessary to get the answer. This is because the first order equation for each $p_i$ must equal the same constant $\beta$, and selecting $p = (1/t) 1$ solves this problem. The function
\[
   \frac{z}{z + x}
\]
is concave in $z$, so $p = (1/t) 1$ is the unique maximum.

\noindent Part \ref{Enum:NEPRisk}. A crowd's risk seeking is underbetting of safe bets and overbetting of long shots (the \emph{favorite-longshot bias}) and a crowd's risk aversion is the reverse. We may express this as follows: let the probabilities $p_1$, $p_2$, \ldots $p_t$ be ranked from highest to lowest 
\[
  p_{(1)}, p_{(2)}, \ldots, p_{(t)}. 
\]
A crowd is \emph{risk seeking} if 
\begin{equation*}
  q_{(i)} = p_{(i)} u_{(i)} \quad \text{for $i = 1, 2, \ldots, t$},
\end{equation*}
where 
\[
  \sum_{i=1}^{i=t} q_{(i)} = 1,
\] 
$u$ is nondecreasing, $u_j \ge 0$, $u_{(1)} < 1$, and $u_{(t)} > 1$. 

\vspace{2mm}
\noindent An analogous definition for \emph{risk averse} requires that $u$ is nonincreasing, $u_j \ge 0$, $u_{(1)} > 1$ and $u_{(t)} < 1$ but such behavior is seldom exhibited at racetracks.

But we are done, since from part (iii) we know that such strategies are asymptotically worse than proportional strategies.

$\blacksquare$

%%************************************************************************************************
\section{Conclusion}

The main result of this paper is that a single syndicate has a mechanical strategy that achieves excess returns against a crowd of uncoordinated bettors when (1) lotteries are equiprobable and have no take (Section \ref{S:EquiprobableLotteriesWithoutCarryoverOrTake}) or (2) lotteries having jackpots and known probabilities satisfy a condition involving the carryover pool, the lottery take and the size of the betting pool. (Section \ref{S:WinningAtNon-UniformLotteries}). 

There are two reasons for these excess returns: convexity of payoffs and a failure to cooperate. Regarding convexity, if payoffs were linear then arguments fail because applications of Jensen's inequality will yield equalities. Regarding non-cooperation, if everyone cooperates then there is one large syndicate that does not have a positive expected return.

Economics explains this sort of behavior by asserting that crowds at racetracks and in lotteries act ``rationally'' using diverse utility functions. And the difficulty of reconciliating those utilities along with a desire (incorporated into them) ``to win the big one'' without sharing with others, promotes  non-cooperation. Of course, this behavior fits nicely with economic theory, but in the process creates an opportunity for substantial returns.

In any case, the rational actor model is difficult to defend as emphasized by Daniel Kahneman:

\small
\begin{quotation}

\noindent ``For emotionally significant events, the size of the probability simply does not matter. What matters is the possibility of winning. People are excited by the image in their mind. The excitement grows with the size of the prize, but it does not diminish with the size of the probability.''  Source: \cite{NYTimes:YourMoney:KahnemanQuote:Online}.

\end{quotation}

Kahneman's behavioral finance explanation leads to a different model of lottery behavior: emotional arousal overwhelms rational, calculated weighing of risks and rewards. The consequence is clear --- ``irrational'' crowd betting will persist, since changing emotional responses is very, very difficult. Moreover, persons motivated by irrational lottery-itis have a behavioral incentive not to join syndicates, because that eliminates the excitement. Therefore, for this inefficiency not to be exploitable, there need to be significant ``limits to arbitrage.'' An example explains one limit to arbitrage in government lotteries:

\vspace{3mm}
\begin{exmp}[\textit{Betting on Lotteries}]\label{Exmp:BettingOnRealLotteries}

Government-sponsored lotteries typically do not want syndicates to play their lotteries, not necessarily because they understand the nature of the edge, but because exposure of a syndicate would lead players to believe that the lottery was ``fixed.'' As a personal experience, one of the authors once approached a lottery asking to bet the entire pool by writing a check (with a bonus to cover the lottery's costs). The lottery manager answered ``If I so much as see any unusual activity at our ticket outlets, I'll cut you off. And the full force of the state police will be brought to bear.'' We took this as a warning not to attempt to cover the pool on our own even though what we wanted to do was perfectly legal.

But apart from such bluster, the logistics of betting $20$ million tickets in a week is considerable. One would have to acquire or reproduce the cards (not difficult, unless on short notice), find a printer that could print them out (also not hard, unless on short notice) and devise a plan to collect tickets from dozens if not hundreds of lottery outlets. And of course, there is the problem of preventing theft of tickets and of arranging payment. 

In short, the logistics of covering a large pool are formidable. %\footnote{One such attempt is rumored, and the syndicate won. But they were lucky; they only had time to bet about $70\%$ according to one source and $85\%$ according to another of all the tickets. \url{http://www.nytimes.com/1992/02/25/us/group-invests-5-million-to-hedge-bets-in-lottery.html?pagewanted=all}. Of course, crowd-sourcing is possible, if issues of trust can be resolved.}

\noindent $\blacksquare$

\end{exmp}

We have demonstrated that there generally exists a purely mechanical strategy that produces excess returns in pure jackpot lotteries. When the lottery has no take and no carryover pool, the edge of a syndicate is reduced, but not eliminated if other syndicates also participate (Section \ref{SS:SyndicateBetss>=t}).  We showed that for lotteries having take $x$, carryover pool $a$, and syndicate and crowd bets of $s$ and $c$, respectively, a single syndicate has an edge if $a/(t+c) - x \ge 0$. But a competing syndicate can convert an apparently favorable bet into an unfavorable one if $a/(2t+c) - x < 0$. 

%Uncoordinated crowd betting in large lotteries leads to uneven coverage of the ticket sample space, and it is this uneven betting along with convexity of payoffs that facilitates syndicate profitability. But if payoffs are linear (as they are in stocks), uneven coverage is not useful unless one can identify underbet and overbet tickets. Thus if stock market investors produce uneven coverages of stocks, then arguments similar to those in this work fail to produce a winning edge.

%But that strategy usually involves large capital outlays and logistical obstacles. Nonetheless, the strategy has been deployed successfully in some government sponsored lotteries.

%************************************************************************************************
\begin{appendices}
\section{$E[1/(1 + X)]$ When $X \sim Bin(c,q)$} \label{App:Eq1-1+X}

We derive the formula
\begin{align}
  E \left[ \frac{1}{1 + X} \right] = \frac{1}{(c + 1) q} \left( 1 - (1 - q)^{c + 1} \right) \label{E:ExpectedValueOf1/(1+X)}
\end{align}
where $X \sim Bin( c , q )$. Using the identity 
\begin{equation*}
  \frac {1} {1 + k}
  \binom{c}{k}
  =
  \frac {1} {c + 1}
  \binom{c+1}{k+1}
\end{equation*}
we calculate
\begin{align*} E \left[ \frac{1}{1 + X} \right]
                  & = \sum_{k=0}^{k=c} 
                         \binom{c}{k}
                         q^{k} \left( 1 - q \right)^{c - k)} \frac{1}{1 + k} \\ 
                  & = \frac{1}{(c+1)q} \sum_{k=0}^{k=c} 
                         \binom{c+1}{k+1}
                          q^{k+1} \left( 1 - q \right)^{c + 1-(k+1)} \\
                  & = \frac{1}{(c + 1) q} \left\{ \left( \sum_{k=0}^{k=c+1} 
                         \binom{c+1}{k}
                         q^{k} \left( 1 - q \right)^{c + 1 - k} \right) 
                       - \left( 1 - q \right)^{c + 1} \right\} \\
                 & = \frac{1}{(c + 1) q} \left( 1 - \left( 1 - q \right)^{c + 1} \right). 
\end{align*}
When $c \gg 1$, the approximations \eqref{E:ApproxE[(1+X)-1]} and \eqref{E:MeanApproxE[(1+X)-1]} will differ little from \eqref{E:ExpectedValueOf1/(1+X)}:
\begin{align} 
  E \left[ \frac{1}{1 + X} \right] &\approx \frac{1}{(c + 1) q} \left( 1 - exp \left( {- (c+1) q} \right) \right) \label{E:ApproxE[(1+X)-1]} \\
                                   &\approx \frac{1}{\mu} \left( 1 - exp \left( {- \mu} \right) \right), \label{E:MeanApproxE[(1+X)-1]}
\end{align}
where $\mu = E[X] = cq$.

%************************************************************************************************
\section{Analysis of Equation \eqref{E:ExpectedReturnFor1/1+k}} \label{S:AnalysisOfEquation1/1+X}

We show that the function 
\begin{equation*}
  f(q) = q^{-1} \left( 1 - \left( 1 - q \right)^{c+1} \right), \label{E:DuplicateOfEquation7}
\end{equation*}
$c \in \mathbb{N}$, $c \ge 2$, is positive, strictly decreasing and strictly convex on $(0,1)$. Using L'H\^{o}spital's rule,  
\[
  \underset{q \rightarrow 0+}{lim} f(q) = \underset{q \rightarrow 0+}{lim} \frac{\frac{d}{dq} \left( 1 - (1 - q)^{c+1} \right) }{\frac{d}{dq}q} = \underset{q \rightarrow 0+}{lim} (c+1) (1 - q)^{c}  = c + 1
\]
and clearly $\underset{q \rightarrow 1-}{lim} f(q) = 1$. If now $f(q)$ is shown to be decreasing, then positivity follows. The first derivative of $f(q)$ is
\begin{align}
  f'(q) & = \frac{ - 1 + (1 - q)^{c + 1} + (c + 1)q(1 - q)^{c} }{q^2} \nonumber \\
        & = - \frac{\sum_{k=2}^{k=c+1} 
                                \binom{c+1}{k}
                              q^k (1 - q)^{c + 1 - k}}{q^2} \nonumber \\
        & = - (c+1)c \sum_{k=0}^{k=c-1} 
                                \binom{c-1}{k}
                               q^k (1 - q)^{c - 1 - k} \frac{1}{(k+1)(k+2)} \label{E:SeriesFormForExpectedValue-1/1+k}\\
        & < 0, \nonumber
\end{align}
where line 2 follows since
\[
 1 = (1 - q + q)^{c+1} = \left( \sum_{k=2}^{k=c+1} 
                           \binom{c+1}{k}
                           q^k (1 - q)^{c + 1 - k} 
                          \right)
                           \, + \, (c + 1) q (1 - q)^{c} \, + \, (1 - q)^{c + 1}
\]
Thus $f(q)$ is strictly decreasing on $(0,1)$. Starting from expression \eqref{E:SeriesFormForExpectedValue-1/1+k}, the second derivative is 
\begin{align}
\begin{split}
  f''(q) & = \frac{d}{dq} \left\{  - (c+1)c \sum_{k=0}^{k=c-1} 
                               \binom{c-1}{k} q^k (1 - q)^{c - 1 - k} \frac{1}{(k+1)(k+2)} \right\} \nonumber \\
         & = - (c+1)c \sum_{k=0}^{k=c-1} 
                               \binom{c-1}{k} k q^{k-1} (1 - q)^{c - 1 - k} \frac{1}{(k+1)(k+2)} \nonumber \\
         & \qquad + (c+1)c \sum_{k=0}^{k=c-1}  
                                \binom{c-1}{k} q^k (c - 1 - k) (1 - q)^{c - 1 - k - 1} \frac{1}{(k+1)(k+2)} \nonumber \\
         & = - (c+1)c(c-1) \sum_{k=0}^{k=c-2}  
                                \binom{c-2}{k} q^{k} (1 - q)^{c - 2 - k} \frac{1}{(k+2)(k+3)} \nonumber \\
         & \qquad + (c+1)c(c - 1) \sum_{k=0}^{k=c-2}  
                                \binom{c-2}{k} q^k (1 - q)^{c - 2 - k} \frac{1}{(k+1)(k+2)} \nonumber \\
         & = (c+1)c(c-1) \sum_{k=0}^{k=c-2}  
                                \binom{c-2}{k}
                               q^{k} (1 - q)^{c - 2 - k} \left( -\frac{1}{(k+2)(k+3)} + \frac{1}{(k+1)(k+2)} \right) \nonumber \\
         & > 0.
\end{split}
\end{align}
for $q \in (0,1)$. Thus \eqref{E:ExpectedReturnFor1/1+k} is convex on $(0,1)$.

%************************************************************************************************
\section{Betting Strategies for Syndicate and Crowd When $s \ge t$}\label{App:BettingStrategiesForSyndicateAndCrowd}

\subsection{Optimal $q$ for the Crowd When the Syndicate Bets $s \ge t$}\label{App:Optimal.q.ForTheCrowd} 

Suppose that the syndicate bets $n$ times on each ticket proportionally to $(1/t) \, 1$, for a total bet of $nt$. If the crowd bets $c$ tickets using probabilities $q = (q_1, q_2, \ldots ,q_t)$, the expected value for the syndicate is
\begin{equation}
  (n t \, + \, c) E\left[ \frac{n}{n + X} \right] = (n t \, + \, c) \sum_{i=1}^{i=t} \frac{1}{t} \sum_{k=0}^{k=c}
          \left( 
            \begin{array} {c}
              c \\
              k \\
            \end{array} 
          \right)
          q_i^k (1 - q_i)^{c-k} \frac{n}{n + k} \label{E:CrowdExpectedq-Gain}
\end{equation}
Probabilities $q_i$ that optimize \eqref{E:CrowdExpectedq-Gain} can be determined from the first order conditions
\begin{equation}
    (n t \, + \, c) \left\{ \frac{c}{t} \sum_{k=0}^{k=c-1} 
        \left( 
          \begin{array} {c}
            c - 1 \\
            k \\
          \end{array} 
        \right)
      q_i^k (1 - q_i)^{c-1-k} \left( \frac{n}{n + k + 1} - \frac{n}{n + k} \right) \right\} - \gamma = 0, \label{E:CrowdExpectedq-GainFirstOrder}
\end{equation}
for a constant $\gamma$ and each $i = 1, 2, \ldots t$. These conditions are met if $q_i = 1/t$ for each $i$. Further, since $n/(n + k)$ is strictly decreasing in $k$, each term in the sum of \eqref{E:CrowdExpectedq-GainFirstOrder} is negative. Differentiating again shows that 
\begin{align*}
    (n t \, + \, c) \left\{ \frac{c(c-1)}{t} \sum_{k=0}^{k=c-1} 
        \left( 
          \begin{array} {c}
            c - 2 \\
            k \\
          \end{array} 
        \right)
      q_i^k (1 - q_i)^{c-1-k} \left( \frac{n}{n + k + 2} \right. \right. \\
    \left. \left. - 2\frac{n}{n + k + 1} \, + \, \frac{n}{n + k} \right) \right\} > 0,
\end{align*}
since $n/(n + k)$ is convex in $k$ and the term in parentheses is the second difference of $n/(n + k)$. It follows that $q_i = 1/t$ minimizes \eqref{E:CrowdExpectedq-Gain} and is the unique solution. Since $q_i = 1/t$ minimizes the expected gain of the syndicate, it is the crowd's optimal choice. 

Using Jensen's inequality and the strict convexity of $n/(n + k)$ with respect to $k$, we have shown that any crowd deviation from $q = e_t$ lowers its expected return.

To prove the general case, consider that the crowd does not know the distribution of bets that the syndicate will use. Therefore, while the syndicate might pick betting sizes $s r_i$ that are not all equal, the crowd cannot know which tickets receive a larger and which a smaller bet. This means that the crowd faces a problam in which the $s r_i$ can be randomized. Thus, payoff terms of the form $n/(n + k$ in \eqref{E:CrowdExpectedq-Gain} will actually be of the form 
\[
  \sum_{j=1}^{j=t} \frac{s r_j}{s r_j + k}
\]
But for fixed $k$ this sum is the same all $q_i$, and since differentating it term by term retains the individual terms' convexity and monotonicity, one is effectively using the argument for fixed $n$.

%************************************************************************************************
\subsection{Optimal Betting for the Syndicate} \label{App:OptimalBettingForTheSyndicate}

If the syndicate bets $n$ of each ticket, then from equation \eqref{E:CrowdExpectedq-Gain} the syndicate's expected gain is 
\begin{equation}
    g(n) = \left\{ (n t \, + \, c) \sum_{i=1}^{i=t} \frac{1}{t} \sum_{k=0}^{k=c}
          \left( 
            \begin{array} {c}
              c \\
              k \\
            \end{array} 
          \right)
          \left( \frac{1}{t} \right)^k \left( 1 - \frac{1}{t} \right)^{c-k} \frac{n}{n + k} \right\} \, - \, nt \label{E:SyndicateExpectedReturnOn-n} 
\end{equation}
Since the function $\phi(x) = a / (a + x)$, $a > 0$ is strictly convex on $[0,\infty)$, and $E[X_i] = c / t$, we get from Jensen's inequality and \eqref{E:SyndicateExpectedReturnOn-n} that 
\begin{equation*}
  g(n) > (n t \, + \, c) \sum_{i=1}^{i=t} \frac{1}{t} \frac{n}{n + c/t}  \, - \, nt = nt - nt = 0.%(n t \, + \, c) \sum_{i=1}^{i=t} \frac{nt}{nt + c} \, - \, nt = 0.
\end{equation*}
Thus the syndicate's expected gain from betting $n$ of each ticket is positive regardless of the amount $c$ bet by the crowd!
 
The inequality of equation \eqref{E:OptimalityOfCrowdBettingForslet} holds only for $0 \le s \le t$. We now determine the best syndicate betting strategy among choices for betting proportions $r = (r_1, r_2, \ldots, r_t)$ for an arbitrary bet size $s > 0$ when fractional bets are possible. The (random) syndicate bet is 
\[
  s \, r_{\!_D}.
\]
where $D$ is a random variable for the winning ticket. Thus the syndicate's gain $G(r,q)$ --- with it understood that the first argument is the probability vector for the syndicate and the second for the crowd --- is
\begin{equation}
  G(r,q) = (s + c) \frac{s \, r_{\!_D} }{s \, r_{\!_D} + X_{\!_D}} - s,
\end{equation}
where $ X_{\!_D} \sim Bin(c,q)$. This coupled with the assumption $q = e_t = t^{-1} 1$ of optimal crowd betting (Appendix \ref{App:Optimal.q.ForTheCrowd}) gives an expected gain of 
\begin{align}
  E[G(r,e_t)] = (s + c) \left( \sum_{i=1}^{i=t} t^{-1} \sum_{k=0}^{k=c} 
                \left( 
                  \begin{array} {c}
                    c \\
                    k \\
                  \end{array} 
                \right)
              t^{-k} (1 - t^{-1})^{c - k} \frac{s \, r_i}{s \, r_i + k} \right) - s. 
              \label{E:SyndicateExpectedGain-r}
\end{align}
for the syndicate. The best proportions $r_i$ for the syndicate should maximize expression \eqref{E:SyndicateExpectedGain-r} and these can be found by maximizing it subject to $\sum_{i=1}^{i=t} r_i = 1$. The first order conditions are 
\begin{equation}
  (s + c) t^{-1} \sum_{k=0}^{k=c} 
                   \left( 
                     \begin{array} {c}
                       c \\
                       k \\
                     \end{array} 
                   \right)t^{-k} (1 - t^{-1})^{c - k} \frac{k}{(s \, r_i + k)^2} \, - \, \gamma = 0, 
                  \label{E:SyndicateBestStrategy-ri}
\end{equation}
where the equations hold for $i = 1, 2, \ldots, t$ and $\gamma$ is a constant. Choosing $r_i = e_t$ satisfies these equations since they are $t$ equations are identical with that choice. Since the function 
\[
  g(x) = \frac{x}{x + k}
\]
is concave in $x$ for each $k$ and therefore the inner sums in \eqref{E:SyndicateExpectedGain-r} are weighted sums of strictly concave functions with positive weights, it follows that \eqref{E:SyndicateExpectedGain-r} is strictly concave in $r$, thus demonstrating $r = e_t = (1/t) 1$ is the unique maximum.

Using this result, we can show that the syndicate's expected gain is always positive if bets are proportional to $e_t$, i.e. in an amount $s e_t$. Using Jensen's inequality, the inner sums of \eqref{E:SyndicateExpectedGain-r} satisfy
\begin{equation*}
 \sum_{k=0}^{k=c} \left( 
                  \begin{array} {c}
                    c \\
                    k \\
                  \end{array} 
                \right) t^{-k} (1 - t^{-1})^{c - k} \frac{s/t}{s/t \, + \, k} > \frac{s/t}{s/t \, + \, c / t} = \frac{s}{s + c}.
\end{equation*}
Therefore, from equation \eqref{E:SyndicateExpectedGain-r}
\begin{align*}
  E[G(e_t,e_t)] & = (s + c) \left( \sum_{i=1}^{i=t} t^{-1} \sum_{k=0}^{k=c} 
                \left( 
                  \begin{array} {c}
                    c \\
                    k \\
                  \end{array} 
                \right)
              t^{-k} (1 - t^{-1})^{c - k} \frac{s/t}{s/t + k} \right) - s. \nonumber \\
            & > (s + c) \left( \sum_{i=1}^{i=t} t^{-1} \frac{s}{s + c} \right) - s \nonumber \\
            & = 0.
\end{align*}
This part shows that the syndicate's expected gain is always positive if fractional bets are possible.

%************************************************************************************************

\end{appendices}

\bibliographystyle{apalike}
\bibliography{AMFWAL}

\begin{thebibliography}{}

\bibitem[Bernard, 2013]{NYTimes:YourMoney:KahnemanQuote:Online}
Bernard, T.~S. (2013).
\newblock Win a lottery jackpot? not much chance of that.
\newblock {\em The New York Times, Your Money}.

\bibitem[Chernoff, 1980]{chernoff1980analysis}
Chernoff, H. (1980).
\newblock {\em An Analysis of the Massachusetts Numbers Game}.
\newblock Technical Report (Massachusetts Institute of Technology. Department
  of Mathematics). Department of Mathematics, Massachusetts Institute of
  Technology.

\bibitem[Ciecka and Epstein, 1996]{970220123219961201}
Ciecka, J. and Epstein, S. (1996).
\newblock State lotteries and externalities to their participants.
\newblock {\em Atlantic Economic Journal}, 24(4):349.

\bibitem[Cook and Clotfelter, 1993]{10.2307/2117538}
Cook, P.~J. and Clotfelter, C.~T. (1993).
\newblock The peculiar scale economies of lotto.
\newblock {\em The American Economic Review}, 83(3):634--643.

\bibitem[Grote and Matheson, 2011]{RePEc:hcx:wpaper:1109}
Grote, K. and Matheson, V. (2011).
\newblock The economics of lotteries: A survey of the literature.
\newblock Working Papers 1109, College of the Holy Cross, Department of
  Economics.

\bibitem[Hausch et~al., 1981]{RePEc:inm:ormnsc:v:27:y:1981:i:12:p:1435-1452}
Hausch, D.~B., Ziemba, W.~T., and Rubinstein, M. (1981).
\newblock Efficiency of the market for racetrack betting.
\newblock {\em Management Science}, 27(12):1435--1452.

\bibitem[MacLean et~al., 1992]{RePEc:inm:ormnsc:v:38:y:1992:i:11:p:1562-1585}
MacLean, L.~C., Ziemba, W.~T., and Blazenko, G. (1992).
\newblock Growth versus security in dynamic investment analysis.
\newblock {\em Management Science}, 38(11):1562--1585.

\bibitem[Myerson, 1997]{myerson1997game}
Myerson, R.~B. (1997).
\newblock {\em Game Theory: Analysis of Conflict}.
\newblock Harvard University Press.

\bibitem[Sauer, 1998]{citeulike:81468}
Sauer, R.~D. (1998).
\newblock {The Economics of Wagering Markets}.
\newblock {\em Journal of Economic Literature}, 36(4):2021--2064.

\bibitem[Thaler and Ziemba, 1988]{citeulike:1337256}
Thaler, R.~H. and Ziemba, W.~T. (1988).
\newblock {Anomalies: Parimutuel Betting Markets: Racetracks and Lotteries}.
\newblock {\em The Journal of Economic Perspectives}, 2(2):161--174.

\bibitem[Ziemba, 2008]{Ziemba2008183}
Ziemba, W.~T. (2008).
\newblock Chapter 10 - efficiency of racing, sports, and lottery betting
  markets.
\newblock In Hausch, D.~B. and Ziemba, W.~T., editors, {\em Handbook of Sports
  and Lottery Markets}, Handbooks in Finance, pages 183 -- 222. Elsevier, San
  Diego.

\bibitem[Ziemba, 2017]{Ziemba:Adventures}
Ziemba, W.~T. (2017).
\newblock {\em The Adventures of a Modern Renaissance Academic in Investing and
  Gambling}.
\newblock World Scientific Publishing Co. Pte. Ltd.

\bibitem[Ziemba, 2018]{Ziemba:ExoticBetting}
Ziemba, W.~T. (2018).
\newblock {\em Exotic Betting at the Racetrack}.
\newblock World Scientific (forthcoming).

\bibitem[Ziemba et~al., 1986]{ziemba1986dr}
Ziemba, W.~T., Brumelle, S.~L., Gautier, A., and Schwartz, S.~L. (1986).
\newblock {\em Dr Z's 6/49 Lotto Guidebook}.
\newblock Dr. Z Investments, San Luis Obispo and Vancouver.

\end{thebibliography}

\end{document}